\documentclass{Interspeech}

\usepackage{acro}
\usepackage{tipa}
\usepackage{booktabs}
\usepackage{multirow}
\usepackage{graphicx}
\usepackage{stfloats}
\usepackage{xurl}

\DeclareAcronym{pttm}{short=PTTM, long={predictive turn-taking model}}
\DeclareAcronym{hri}{short=HRI, long={human-robot interaction}}
\DeclareAcronym{snr}{short=SNR, long={signal-to-noise ratio}}
\DeclareAcronym{asr}{short=ASR, long={automatic speech recognition}}
\DeclareAcronym{avsr}{short=AVSR, long={audio-visual speech recognition}}
\DeclareAcronym{vap}{short=VAP, long={voice-activity projection}}
\DeclareAcronym{mmvap}{short=MM-VAP, long={multimodal VAP}}
\DeclareAcronym{h/s}{short=H/S, long={hold/shift}}

\interspeechcameraready

\title{Visual Cues Support Robust Turn-taking Prediction in Noise}

\author[affiliation={}]{Sam}{O'Connor Russell}
\author[affiliation={}]{Naomi}{Harte}

\affiliation[nocounter]{ADAPT Centre, School of Engineering}{Trinity College Dublin}{Ireland}
\email{russelsa@tcd.ie, nharte@tcd.ie}

\usepackage{comment}

\begin{document}

\maketitle

\begin{abstract}
        Accurate predictive turn-taking models (PTTMs) are essential for naturalistic human-robot interaction. However, little is known about their performance in noise. This study therefore explores PTTM performance in types of noise likely to be encountered once deployed. Our analyses reveal PTTMs are highly sensitive to noise. Hold/shift accuracy drops from 84\% in clean speech to just 52\% in 10 dB music noise. Training with noisy data enables a multimodal PTTM, which includes visual features to better exploit visual cues, with 72\% accuracy in 10 dB music noise. The multimodal PTTM outperforms the audio-only PTTM across all noise types and SNRs, highlighting its ability to exploit visual cues; however, this does not always generalise to new types of noise. Analysis also reveals that successful training relies on accurate transcription, limiting the use of ASR-derived transcriptions to clean conditions. We make code publicly available for future research. 

\end{abstract}

\section{Introduction}
\label{sec:intro}

Human turn-taking is a remarkable process. There is only 200 milliseconds of silence on average between speaking turns \cite{stivers2009universals}, yet language production takes over 600 milliseconds \cite{indefrey2011spatial}. Listeners must therefore plan what they are about to say whilst their conversational partner is still speaking \cite{sacks1974simplest}. Although effortless for humans, turn-taking is highly challenging for robots. Most turn-taking algorithms in use today are based on a robot reacting to silence after a turn \cite{skantze2021turn}, resulting in interactions that are less fluid than those between humans \cite{li2022can,woodruff2003push}. \Acp{pttm} have been proposed to overcome these limitations \cite{skantze2017towards,skantze2021turn}. \Acp{pttm} learn human-like turn-taking from large corpora of human interaction, e.g. continuously predicting whether a shift between speakers will occur \cite{skantze2017towards,roddy2018investigating,ekstedt2022much,li2022can}, akin to human turn-taking \cite{sacks1974simplest}. 

Most \acp{pttm} use speech features \cite{skantze2021turn}. Yet in human-human interaction, listeners make faster and more accurate turn-taking decisions when they can see and hear a speaker \cite{barkhuysen2008interplay}. Recent work shows multimodal \acp{pttm}, which also use gaze, head pose and facial expression, outperform audio-only \acp{pttm} \cite{onishi2023multimodal,russellACL}, illustrating the known role of visual cues in turn-taking \cite{holler2016turn,barkhuysen2008interplay}. This reflects the many communicative advantages of multimodality \cite{holler2019multimodal}, which include boosting the intelligibility of speech in noise \cite{macleod1987quantifying}. MacLeod and Summerfield found that the intelligibility gain from being able to see a speaker was equivalent to an 11 dB increase in the \ac{snr} \cite{macleod1987quantifying}. This motivates \ac{avsr}, which exploits visual cues to overcome the performance degradation of \ac{asr} in noise \cite{shi2022robust}. 

This paper asks if multimodality confers similar performance benefits to \acp{pttm} in noisy environments. Surprisingly, \acp{pttm} have to date only been tested on interactions free from background noise interference \cite{skantze2021turn,onishi2023multimodal}. The impact of noise on \ac{pttm} performance is therefore unknown and unquantified. This presents a major oversight as \acp{pttm} will be deployed \textit{in-the-wild}, where they will inevitably encounter diverse sources and levels of noise. We therefore present the first exploration of PTTM performance in background noise, asking:

% \subsection{Research objectives and contributions}

\begin{enumerate}
    \item \textit{How is the performance of predictive turn-taking models affected by background noise?} and
    \item \textit{Does the inclusion of visual features make multimodal \acp{pttm} robust to background noise?}
\end{enumerate}

This paper explores the crucial capability of a \ac{pttm} to distinguish holds (no speaker change) from shifts (changes between speakers). In noise-free conditions, a multimodal \ac{pttm} which incorporates visual features outperforms an audio-only model ($84\%$ and $80\%$ accuracy, Table \ref{tab:tabluated_accuracy}). When we add artificial music, babble, and speech noise, the accuracy of both \acp{pttm} is poor ($51$\% and $52\%$ in music). Including noise during training increases accuracy ($61$\% and $72\%$ in music). It also \lq primes' the multimodal \ac{pttm} to better exploit visual information (10 dB effective SNR gain over the audio-only \ac{pttm}). We find this effect generalises to types of noises unseen by the model during training. The inclusion of visual cues therefore makes PTTM performance more robust to noise. We also investigate the reliance of \acp{pttm} training on accurate transcriptions. When noisy data is re-transcribed with ASR PTTM performance collapses ($50$-$54\%$). \acp{pttm} training therefore relies heavily on accurately aligned transcripts collected from noise-free recordings, complicating the use of ASR in noisy data. 

PTTMs are essential to facilitating interaction between humans and robots, but there is little understanding of how well they could work in-the-wild. Part of the puzzle is understanding how they handle noise and our work therefore represents a significant step forward in \acp{pttm} development. 

\section{Predictive turn-taking models}
\label{sec:pttm-model-outline}

\Acp{pttm} continually predict future speaking activity, from which turn-taking predictions can be made \cite{skantze2017towards,roddy2018investigating,ekstedt22_interspeech}. For example, if speaker 0 is currently talking, but the model prediction favours speaker 1, a \textit{shift} is predicted. \Acp{pttm} are developed on corpora of human-human interaction with the long-term eventual goal of deployment to human-robot interaction. Our work considers two-party interaction, where most work has been conducted \cite{skantze2021turn}. 

We evaluate two \acp{pttm} based on the state-of-the-art \textbf{\acf{vap}} approach introduced by Ekstedt et al. \cite{ekstedt22_interspeech} as it has been thoroughly explored \cite{ekstedt23_interspeech,ekstedt2022much,onishi2023multimodal}. \Ac{vap} models predict future speaking activity in 8 bins (4 per speaker). At time $t=0$, bins span $[0,0.2], [0.2,0.6], [0.6,1.2]$ and $[1.2,2.0]$ seconds into the future. Each bin has a binary label (1=speech if $>50\%$ frames contain speech), giving 256 possible outputs. 

\Ac{vap} models are neural networks consisting of stacked transformer decoder layers \cite{ekstedt22_interspeech,inoue-etal-2024-multilingual}. The first model is an \textbf{audio-only VAP model} which we re-implement from \cite{inoue-etal-2024-multilingual}. The initial \ac{vap} model used mono audio and ground-truth knowledge of the current speaker \cite{ekstedt22_interspeech}; i.e. it knew when each turn ended. The updated model used here takes stereo audio with 1 channel per speaker and removes the dependency on ground-truth knowledge, reflecting a real-world deployment. We also use a \textbf{\acf{mmvap}} model that incorporates facial action unit, facial landmark, gaze and head pose features from OpenFace \cite{baltruvsaitis2016openface,onishi2023multimodal,russellACL}. As the initial \ac{mmvap} model also depended on the ground-truth knowledge of the current speaker \cite{onishi2023multimodal}, we similarly extended the model to support stereo audio and remove this dependency \cite{russellACL}. For comparison, we also train a \textbf{video-only \ac{vap}} model, identical to \ac{vap} with the audio encoder removed \cite{inoue-etal-2024-multilingual}. Our implementations, along with full architectural details are publicly available in our repository\footnote{\url{https://github.com/russelsa/mm-vap}}.

\Acp{pttm} require labels of the start and end of each utterance for training (the \textbf{\textit{alignment}}). Manual transcriptions are typically used e.g. the Switchboard telephone corpus \cite{godfrey1992switchboard,ekstedt22_interspeech,roddy-harte-2020-neural,li2022can}. \Ac{asr} more closely represents real-world deployments and has been shown to work well in \ac{pttm} training \cite{russell2024automatic,russell2024towards}. 

\section{Data preparation} 
\label{sec:dataprep}
% \subsection{The Candor corpus of human interaction}
This work uses the Candor corpus \cite{reece2023candor}, an 850 hr corpus of 1657 unscripted two-party videoconferencing interactions between speakers in US English. The average session duration is 34 min. Recordings are 60 fps video, 32 kHz stereo audio with one channel per speaker. There is no controlled background noise. Recordings take place at a location of the participant's choosing e.g. home offices. We transcribe with Amazon Transcribe (EN-US model) which outputs word-level timings and preserves most disfluencies \cite{russell2024automatic}. We extract gaze, head pose, facial action units, and facial landmarks with OpenFace \cite{baltruvsaitis2016openface}. 

We follow the \ac{avsr} literature by artificially simulating noisy conditions, adding noise separately to each channel at controlled \acp{snr} \cite{shi2022robust}. \textbf{Babble noise} consists of 65,000 unique 30 speaker \textbf{babbles} which is synthesised from LRS-3 \cite{afouras2018lrs3,shi2022robust} (46 hr train, 14 hr test). \textbf{Music noise} is sourced from MUSAN \cite{snyder2015musan} (15 hr train from the \textit{Jamendo} genre, 70 hr test). Interference from other speakers (\textbf{speech noise}) is simulated by overlaying another speaker from Candor, ensuring train/test separation. 

In \ac{avsr} research, noise is typically added to isolated speech segments. However, \acp{pttm} are trained using conversational speech which includes long pauses. A standard method of adding controlled noise to conversational speech has not been reported. We propose to add noise at a given SNR by considering signal energy during speaking times and then adding scaled noise to the complete recording. Natural variation in amplitude means this yields an average SNR over all utterances.

The \textbf{clean test audio} comprises 70 sessions withheld for evaluation. The test audio is also corrupted with -10 to +10 dB \acp{snr} noise in 2.5 dB increments (\textbf{noisy test audio}). Shi et al. found that adding noise in training increased \ac{avsr} performance. Following their method, in each training session, with probability 0.25 we add either babble, music or speech noise at 0 dB \cite{shi2022robust}, yielding the \textbf{augmented training audio}. Note that clean and augmented training sets are the same length. We obtain both \textbf{clean} and \textbf{augmented alignments} for both training sets with Amazon ASR. 

\section{Model training and evaluation}
\label{sec:model-eval}
First, we train with the clean audio + clean alignment, as is standard in the turn-taking literature \cite{ekstedt22_interspeech,russell2024towards}. Then we train on the augmented audio + clean alignment, where noise is added but the transcription remains unchanged \cite{shi2022robust}. To test the reliance on accurate ASR, we train using the augmented audio + augmented alignment. This represents a situation where there is noisy data transcribed with ASR. Training batches consist of 20-second windows with 2 second overlap randomly sampled from training sessions. The training procedure is a 5-fold cross-validation with 10 epochs per fold on an NVidia RTX 6000 GPU. A grid search sets the batch size to 16 and the learning rate to 1e-4. The loss function is the sum of cross-entropy (future speaking prediction) and binary cross-entropy (current speaker prediction) losses minimised with AdamW \cite{loshchilov2017decoupled}. 

We report the \ac{h/s} prediction accuracy: the ability of the model to distinguish holds (no speaker change) from shifts (speaker change), a common metric in the turn-taking literature \cite{ekstedt22_interspeech,inoue-etal-2024-multilingual}. First, all silences $> 250$ ms are identified. Those with speaker changes are shifts (83,158) and those without are holds (206,830) \cite{ekstedt22_interspeech}. Then, the probability of speech in the latter half of the \ac{vap} bins is computed \cite{ekstedt22_interspeech}. If speaker 0 has finished speaking and the probability of speaker 1 exceeds a binary threshold, a shift is predicted. We set the threshold to maximise balanced accuracy on validation sessions and report a 5-fold average on test sessions. 

We report effective SNR gain i.e. the SNR at which \ac{mmvap} achieves the same accuracy as \ac{vap} at 0 dB, following the \ac{avsr} literature \cite{lin2025uncovering}. We compare \ac{vap} and \ac{mmvap} model performance with two-tailed independent t-tests \cite{student1908probable}. 

\section{Turn-taking model performance in noise}
\label{sec:results}

% \usepackage{booktabs}
% \usepackage{multirow}
% \usepackage{graphicx}
\begin{table*}[!t]
\centering
\caption{Average hold/shift prediction accuracy (\%, $\bar{x}$=mean all SNRs) across 5-folds of cross-validation of VAP (audio, A), MM-VAP (audio+video, A+V), and video-only (V) models on the same test set at varying noise levels (2.5 dB increments, 5 dB shown for brevity). Training is conducted on the clean and augmented (aug., 25\% of sessions corrupted with 0 dB noise) audio and ASR alignments. 
}
\label{tab:tabluated_accuracy}
\resizebox{\textwidth}{!}{%
\begin{tabular}{@{}llll|rrrrrrrrrrrrrrrrrr|l@{}}
\toprule
  \multicolumn{1}{c}{\multirow{1}{*}{\textbf{Model}}} &
  \multicolumn{2}{c}{\multirow{1}{*}{\textbf{Training}}} &
  % \multicolumn{1}{c}{\multirow{2}{*}{\textbf{Labels}}} &
  \multicolumn{1}{c|}{\multirow{1}{*}{\textbf{Input Modality}}} &
  \multicolumn{6}{c|}{\textbf{Test set + babble @} \textit{\textbf{SNR dB =}}} &
  \multicolumn{6}{c|}{\textbf{Test set + speech @} \textit{\textbf{SNR dB =}}} &
  \multicolumn{6}{c|}{\textbf{Test set + music @} \textit{\textbf{SNR dB =}}} &
  \textbf{Clean} \\
  \multicolumn{1}{c}{} &
  \multicolumn{1}{c}{\textbf{\textit{Audio}}} &
  \multicolumn{1}{c}{\textbf{\textit{Alignment}}} &
  \multicolumn{1}{c|}{} &
  \textit{-10} &
  \textit{-5} &
  \textit{0} &
  \textit{5} &
  \textit{10} &
  \multicolumn{1}{r|}{$\bar{x}$} &
  \textit{-10} &
  \textit{-5} &
  \textit{0} &
  \textit{5} &
  \textit{10} &
  \multicolumn{1}{r|}{$\bar{x}$} &
  \textit{-10} &
  \textit{-5} &
  \textit{0} &
  \textit{5} &
  \textit{10} &
  \multicolumn{1}{r|}{$\bar{x}$} &
  $\infty$ \\ \midrule
\textit{VAP} & clean & clean & A & 61 & 50 & 58 & 69 & 76 & \multicolumn{1}{r|}{60} & 51 & 51 & 51 & 51 & 52 & \multicolumn{1}{r|}{51} & 51 & 50 & 51 & 51 & 52 & 51 & 80 \\
\textit{MM-VAP} & clean & clean & A+V & 57 & 56 & 68 & 76 & 80 & \multicolumn{1}{r|}{68} & 53 & 52 & 52 & 53 & 54 & \multicolumn{1}{r|}{52} & 53 & 51 & 51 & 52 & 53 & 52 & 84 \\ \midrule
\textit{VAP} & aug. & clean & A & 49 & 50 & 65 & 74 & 77 & \multicolumn{1}{r|}{63} & 51 & 59 & 65 & 66 & 64 & \multicolumn{1}{r|}{61} & 55 & 58 & 62 & 66 & 65 & 61 & 84 \\
\textit{MM-VAP} & aug. & clean & A+V & 61 & 62 & 75 & 79 & 80 & \multicolumn{1}{r|}{71} & 66 & 69 & 74 & 74 & 75 & \multicolumn{1}{r|}{72} & 67 & 68 & 74 & 76 & 74 & 72 & 86 \\ \midrule
\textit{VAP} & aug. & aug. & A & 50 & 50 & 55 & 55 & 56 & \multicolumn{1}{r|}{53} & 50 & 50 & 50 & 49 & 49 & \multicolumn{1}{r|}{50} & 50 & 50 & 50 & 50 & 50 & 50 & 57 \\
\textit{MM-VAP} & aug. & aug. & A+V & 55 & 56 & 57 & 58 & 58 & \multicolumn{1}{r|}{57} & 50 & 50 & 50 & 50 & 49 & \multicolumn{1}{r|}{50} & 50 & 50 & 50 & 50 & 50 & 50 & 57 \\
\midrule
\textit{video-only} &
   \multicolumn{1}{c}{---} &
   \multicolumn{1}{l}{clean} &
   \multicolumn{1}{l|}{V} &
   \multicolumn{6}{c|}{66} &
   \multicolumn{6}{c|}{66} &
   \multicolumn{6}{c|}{66} &
   \multicolumn{1}{l}{66} \\\bottomrule
\end{tabular}%
}
\end{table*}

\begin{figure*}[]
    \centering
    \begin{minipage}{0.48\textwidth}
        \centering
        \includegraphics[width=0.865\columnwidth]{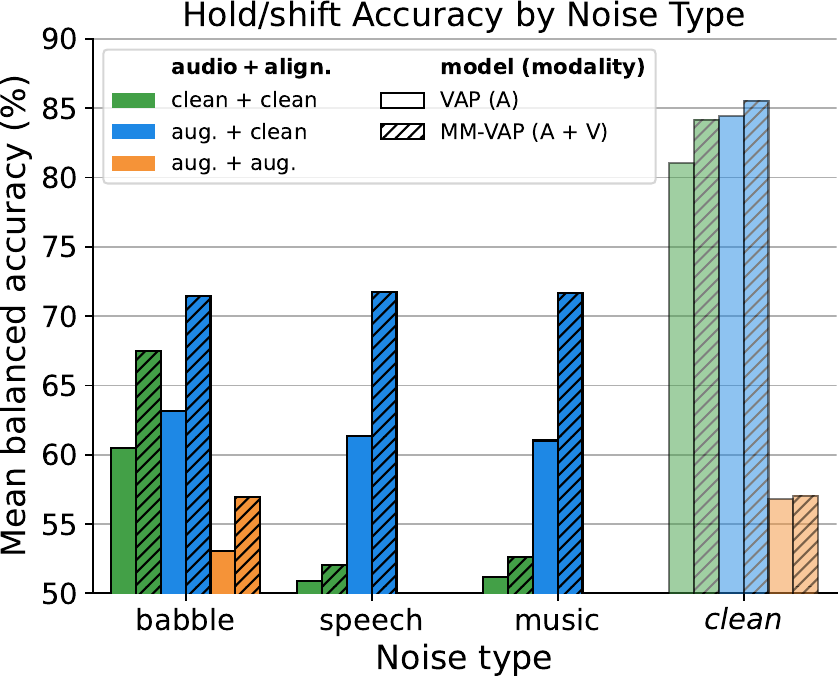}
        \caption{Average hold/shift prediction accuracy of VAP (audio-only) and MM-VAP (audio+video) models in each noise type. The average is across the -10 dB, +10 dB SNR range.}
        \label{fig:bar_noise_type}
    \end{minipage}
    \hfill
    \begin{minipage}{0.48\textwidth}
        \centering
        \includegraphics[width=0.70\columnwidth]{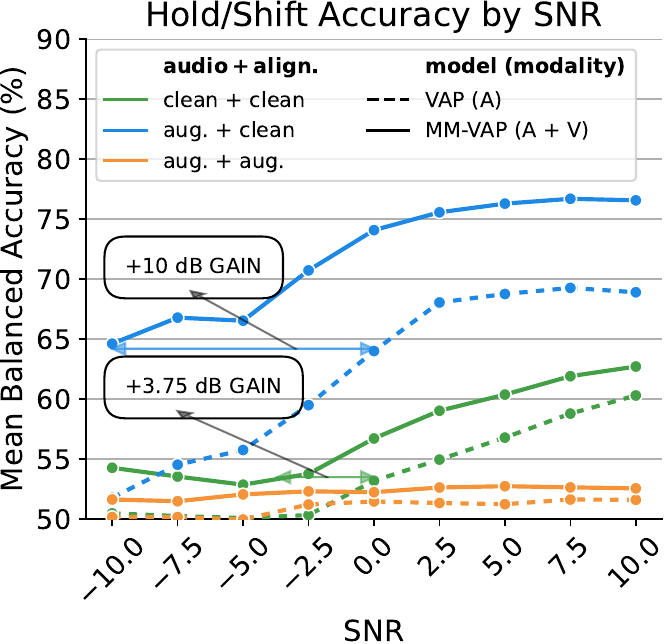}
        \caption{Average hold/shift prediction accuracy of VAP (audio-only) and MM-VAP (audio+video) models at each SNR. An average accuracy in speech, music and babble noise is shown.}
        \label{fig:line_snr}
    \end{minipage}
    % \caption{Overall caption for both figures}

\end{figure*}

\subsection{Baseline performance on clean speech}
\label{subsec:cleanspeechbaseline}
We begin our analysis by training \ac{vap} and \ac{mmvap} models using the clean audio and clean alignment, establishing a baseline \ac{h/s} prediction accuracy without added noise (clean test audio, Table \ref{tab:tabluated_accuracy}, first two rows). The 5-fold average \ac{h/s} prediction accuracy of \ac{mmvap} model is 84\%, greater than the audio-only \ac{vap} at 80\% ($p<0.001$). The accuracy and the superior performance of \ac{mmvap} are in agreement with the \ac{pttm} literature \cite{inoue-etal-2024-multilingual,russell2024towards}. 

\subsection{The impact of noise on turn-taking models}
\label{subsec:cleantrainedmodels}
Figure \ref{fig:bar_noise_type} shows the average \ac{h/s} accuracy on the noisy test data, where solid colours are \ac{vap} (audio-only) and hatched are \ac{mmvap} (audio and video). We first consider the models trained on clean audio + alignment in green. The plot reveals noise has a strong impact on PTTM accuracy, most severe in speech and music where accuracy falls to $51$-$52\%$ (the first two rows of Table \ref{tab:tabluated_accuracy}). At high SNR there is a smaller impact from babble noise at $76$-$80\%$ at 10 dB SNR. However, both models are extremely sensitive to music and speech interference with $51$-$52\%$ accuracy at 10 dB \ac{snr}, i.e. a random hold/shift prediction.

 % Again,  models trained on the clean audio and alignment are plotted in green. 
Figure \ref{fig:line_snr} shows average \ac{h/s} prediction accuracy at each \ac{snr} for \ac{vap} and \ac{mmvap}. The higher the SNR, the lower the noise energy relative to the speech signal (equal at 0 dB). Above 0 dB, the average accuracy increases as the SNR increases. Below 0 dB, accuracy levels off to 50-54\%. The effective SNR gain of \ac{mmvap} over \ac{vap} is +3.75 dB. A breakdown of average accuracy by noise type reveals the superior performance of \ac{mmvap} originates from accuracy in babble noise, as accuracy in music and speech is always $51$-$54\%$ (Table \ref{tab:tabluated_accuracy}). 

\subsection{Training with noisy examples}
\label{subsec:augtrainedmodels}
We re-train both models on the augmented training set and show \ac{h/s} prediction accuracy by noise type in blue in Figure \ref{fig:bar_noise_type}. There is a clear performance increase, e.g. average accuracy in music is $61\%$ for \ac{vap} and $72\%$ for \ac{mmvap}, up from $51$-$52\%$ ($p<0.001$). A large gap in model accuracy has also emerged. 

The blue lines in Figure \ref{fig:line_snr} show performance at each SNR. At -10 dB SNR, the average accuracy of \ac{vap} is 50\% whereas \ac{mmvap} is 65\% ($p<0.01$). The effective SNR gain of \ac{mmvap} has increased to +10 dB and the extreme sensitivity to speech and music noise is reduced. At +10 dB SNR of speech noise, the average accuracy is $64\%$ for \ac{vap} and $75$\% for \ac{mmvap} (second two rows of Table \ref{tab:tabluated_accuracy}), up from $51$\% and $52$\% previously ($p<0.01$). At -10 dB SNR, \ac{vap} has a 50\% accuracy. On the other hand, \ac{mmvap} backs off to the accuracy of a \ac{pttm} model trained exclusively on video features (67\% vs 66\% in -10 dB music noise). The only difference in input is \ac{mmvap} receives visual features. However, it achieves superior performance in all types of noise and across the full range of \acp{snr}, and performance is similar to a video-only model in extreme (-10 dB) noise. This demonstrates that it exploits visual information to improve accuracy in acoustic interference. 

\subsection{Predicting a shift between speakers}
\label{subsec:shiftpredexample}
\begin{figure}[]
    \centering
    \includegraphics[width=\columnwidth]{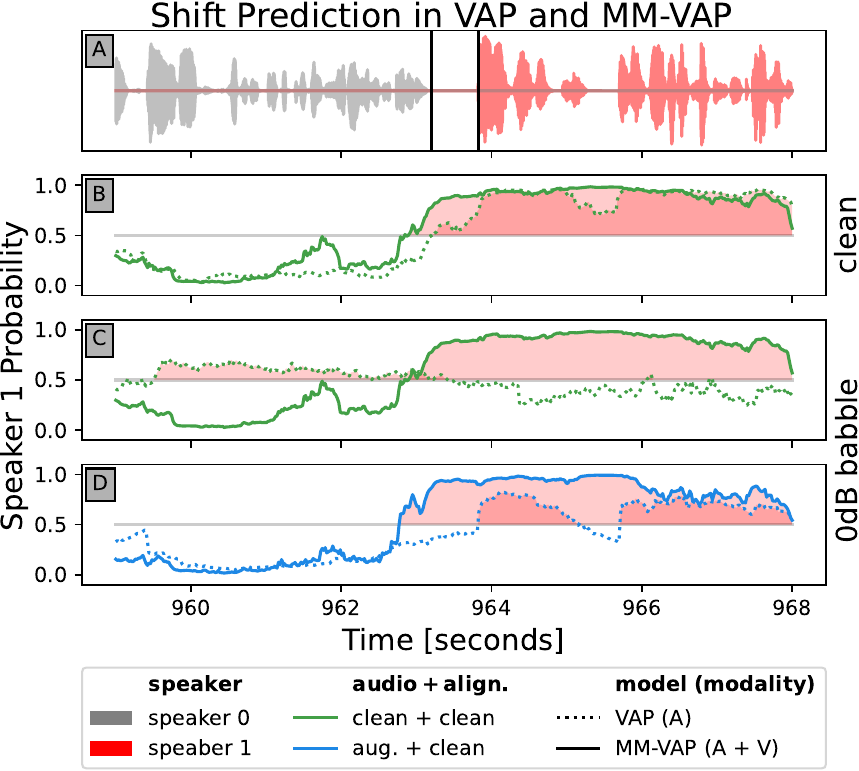}
    \caption{Model output during a shift in the Candor corpus (unseen during training). Speaker 0 (grey): \textit{"So uhm would you feel comfortable telling me about what you do for a living?"}, Speaker 1 (red) \textit{"Uhm yeah well for the most part I test and tweak algorithms"}. VAP and MM-VAP model output is shown in clean speech (B) and 0 dB babble speech (C) and (D). }
    \label{fig:sample-predictions}
\end{figure}

We show predictions during a shift in Figure \ref{fig:sample-predictions} (\ac{mmvap} solid, \ac{vap} dashed) by plotting speaker 1 probability in the next 1.2-2 seconds (greater than $0.5$ highlighted red). In clean audio (B), the shift between speakers is predicted ahead of time by \ac{mmvap}. In 0 dB babble noise (C), \ac{mmvap} predicts the shift, whereas \ac{vap} always outputs approx 0.5, demonstrating the impact of noise. \Ac{mmvap} predictions in noise (C + D) resemble predictions in clean audio (B), further demonstrating the exploitation of visual cues by \ac{mmvap}. 

\begin{table}[]
\centering
\caption{\Ac{h/s} prediction accuracy (\%) of MM-VAP trained with speech, babble, music, or a mixture (augmented) at 0 dB SNR.}
\label{tab:model-comp}
\resizebox{0.85\columnwidth}{!}{%
\begin{tabular}{llrrrr}
\hline
 &  & \multicolumn{4}{c}{\textbf{Test set, 0 dB SNR}} \\
 &  & \multicolumn{1}{l}{\textit{babble}} & \multicolumn{1}{l}{\textit{music}} & \multicolumn{1}{l}{\textit{speech}} & \multicolumn{1}{l}{\textit{clean}} \\ \hline
\multirow{4}{*}{\rotatebox[origin=c]{90}{\textbf{Train set}}} & \textit{babble} & 71 & 51 & 52 & 85 \\
 & \textit{music} & 58 & 75 & 57 & 86 \\
 & \textit{speech} & 70 & 77 & 76 & 86 \\
 & \textit{augmented} & 75 & 74 & 74 & 86 \\ \cline{1-6} 
\end{tabular}%
}
\end{table}

\subsection{Generalisation to noise unseen during training}
\label{subsec:generalisation}
We investigate if \ac{mmvap} performance extends to types of noise unseen during training. We retrain three times with either 0 dB SNR speech, music or babble noise in 25\% of training sessions. The balanced accuracy of \ac{h/s} prediction in Table \ref{tab:model-comp} shows that when the training set includes speech or babble noise, accuracy is poor in other types of noise: e.g. $71\%$ for the babble-trained model in babble noise but only $54\%$ in music. However, when trained on speech noise, similar accuracy is obtained in all types of noise $74-75\%$ (Table \ref{tab:model-comp}, second-last row). This suggests training on a speech signal with an interfering speaker \lq forces' \ac{mmvap} to exploit visual cues, leading to better generalisation when new types of noise are encountered. 

\subsection{The critical role of the alignment in training}
\label{subsec:alingmentimpact}
Finally, we re-train both models using the augmented alignment obtained from Amazon ASR. The orange bars in Figure \ref{fig:bar_noise_type} reveal a stark fall in accuracy to $50-58\%$ and Figure \ref{fig:line_snr} shows that this is consistent across all \acp{snr}. The drop in performance to the 50\% range indicates models are comparable to a random hold/shift prediction. This is due to alignment and transcription errors that ASR experiences in noise \cite{marxer2018impact}. PTTM training is therefore dependent on accurate transcriptions which are either manual \cite{ekstedt22_interspeech} or obtained from ASR and clean recordings \cite{russell2024towards}. 

 % and won't yield an accurate transcription
\section{Discussion}

% \subsection{Turn-taking models show limited robustness to noise}
Our results show \acp{pttm} do not maintain high accuracy in types of noises they are likely to encounter in-the-wild (Section \ref{subsec:cleantrainedmodels}). We found slight (10 dB SNR) interference from music and other speakers caused accuracy to plummet to 50\%. Including noise during training dramatically increased performance (Section \ref{subsec:augtrainedmodels}). This is a common technique in \ac{avsr} \cite{shi2022robust} that has not previously been considered in the turn-taking literature. Returning to our first research question, our findings therefore show that the \acp{pttm} in this work are not robust to noise without the prior inclusion of noise in training. 

% \subsection{Visual information can improve turn-taking in noise}
% concerned whether visual information makes multimodal \acp{pttm} more robust to noise. 

The initial poor performance shows \ac{mmvap} does not exploit visual cues to their fullest extent (Section \ref{subsec:cleantrainedmodels}) without appropriate training. Training on augmented data led to considerably more improvement in the accuracy of \ac{mmvap} than \ac{vap} across the SNR range. At extreme levels of noise interference (e.g. -10 dB music noise) \ac{vap} fails (50\%, i.e. chance) but \ac{mmvap} backs off to the accuracy of the video-only model (67\% vs 66\%, Section \ref{subsec:augtrainedmodels}). This highlights the role visual cues play in achieving robust H/S accuracy in noise. 

If \ac{mmvap} is truly able to utilise visual cues to overcome noise, this should generalise to new sources of noise. However, we found that this was only true for the \ac{mmvap} model trained on speech-noise (Section \ref{subsec:generalisation}). Our working hypothesis is that mixing in random speakers over the speech signal training process \lq forces' \ac{mmvap} to use visual information, rather than relying on the acoustic properties of the noise. Thus, returning to our second research question, \acp{pttm} can achieve robust performance in noise, but this is intimately related to the training process and does not inherently generalise to new noise types. 

It is difficult to interpret \ac{h/s} prediction accuracy numbers as no studies have determined a minimum accuracy needed to support fluid interaction in a human-robot deployment. Taking our 80-84\% accuracy on clean speech as a baseline and ignoring the many complexities of a real-world deployment including hardware, real-time performance and human-machine dialogue, a robot would make a correct \ac{h/s} decision 8/10 times. Both \ac{mmvap} and \ac{vap} would only be correct 5/10 times in 10 dB SNR music interference when trained using conventional methods, significantly complicating interaction. This rises to 6.5/10 for \ac{vap} and 7.5/10 for \ac{mmvap} when noise is included in training; close to baseline performance, illustrating the power of visual cues in achieving robust performance in noise.

% \subsection{Limitations and future work}
A limitation of our work is we artificially add noise to speech recorded in noise-free settings. Although this is commonplace \cite{marxer2018impact,shi2022robust,lin2025uncovering}, it means we do not capture changes to verbal and non-verbal \cite{trujillo2021speakers} communication and turn-taking \cite{sorensen2020effects} in noise. Another limitation is our use of videoconferencing. However, no in-person corpora match the size and quality of Candor. Finally, our mixing of noise at a constant energy level means the audio-only \ac{vap} model may be partly capitalising on the natural variations in SNR in long recordings of speech. 

Training techniques for multimodal \acp{pttm} should be further explored to ensure their ability to exploit visual cues generalises to new noise types. As AVSR can exploit visual information in datasets unseen in training \cite{sterpu2020teach}, AVSR forms a natural starting point for future multimodal \acp{pttm} architectures \cite{shi2022robust}. Turn-taking in noise is quite similar to the channel separation problem of speaker diarisation  \cite{chen24d_interspeech} and our findings support further exploration and the extension to the multimodal setting. There are many other turn-taking events (e.g. backchannels) which require consideration in future work. 

\section{Conclusion}

Accurate turn-taking models are an important milestone on the path to viable human-robot interaction. We have conducted the first exploration of \ac{pttm} performance in noise, showing that \acp{pttm} are not robust to typical noises a \ac{pttm} may encounter once deployed. Training with noise improves overall accuracy and enables multimodal \ac{pttm} to leverage visual features to overcome acoustic interference. However, this capability does not necessarily generalise to new types of noise. This issue must be solved prior to their deployment in real-world settings. Our work therefore constitutes a significant step forward in our understanding of \acp{pttm}. Future work in \ac{pttm} architectures, training techniques and in-the-wild deployment are therefore imperative before deployment to human-robot interaction. 

\section{Acknowledgements}
This research was conducted with the financial support of Science Foundation Ireland under Grant Agreement No. 13/RC/2106\_P2 at the ADAPT SFI Research Centre at Trinity College Dublin. Amazon provided academic access to the AWS Transcribe service. 

\bibliographystyle{IEEEtran}
\bibliography{main}

\end{document}